\documentstyle[aps,epsfig]{revtex}
\newcommand{\dis}{\displaystyle}
\newcommand{\hyp}{{}_2 F_{1}}
\newcommand{\sde}{Schwinger-Dyson equation}
\newcommand{\dirac}{\not\!}
\newcommand{\myGamma}[2]{\Gamma \left[
\begin{array}{c} #1 \\ #2 \end{array} \right] }
\newcommand{\MS}{$\overline{MS}$}
\begin{document}

\title{Light-cone representation of the quark \sde.}

\author{L. S. Kisslinger and O. Linsua\'{\i}n}
\address{Department of Physics,\\
      Carnegie Mellon University, Pittsburgh, PA 15213} 

\maketitle
\begin{abstract}

We use a light-cone approach to solve the \sde\ for the quark propagator.
We show how this method can be used to solve the equation beyond the
space-like region, to which one is usually restricted with the
Euclidean-space approach. We work in the Landau gauge, and use an 
infrared-enhanced model for the gluon propagator and include instanton
effects to get both confinement and vacuum condensates. With our models 
reasonable fits to known quantities are obtained, resulting in a light-cone
quark propagator that can be used for hadronic physics at all momentum
transfers.

\end{abstract}

\section{Introduction.}\label{intro}

\hspace{3mm}
   For a microscopic QCD description of hadrons and hadronic matter one
needs the fully dressed nonperturbative quark and gluon propagators, for
which the Schwinger-Dyson formalism is a natural approach. 
A full study of QCD, however, requires investigation of  hadronic 
properties at all momentum transfers. Since instant form of field 
theory is difficult to use for composite states at medium or high 
momentum, a light-cone  representation is desirable\cite{dirac}. 
In the present paper we develop a light-cone
formulation of the Schwinger-Dyson equation for the quark propagator
for use in developing hadronic light-cone Bethe-Salpeter amplitudes
as well as providing new aspects of the quark propagator, which we discuss 
below.

It is well known that short distance processes occurring inside hadrons are
well described by perturbative QCD (p-QCD) calculations. It is also well 
known that quarks never emerge as such from the reaction region, but 
instead they hadronize at a distance of about a fermi. But often these
processes are not well described by p-QCD. The reason is that perturbative 
calculations in QCD have a rather limited range of validity: they are a 
good approximation only at very short distances where the expansion 
parameter (the effective QCD coupling constant)
is small. But if QCD is the correct theory of the strong interactions (and
that is the consensus today), then it should describe medium and long 
distance hadronic processes as well. We need then to understand how to 
compute the basic entities of the theory (its Green's functions) not just 
over very short distances, but over long distances too.

The Schwinger-Dyson equations of a field theory embody all its 
dynamics~\cite[p. 475]{zuber}. They are the complete equations of motion for the 
Green's functions of the theory, and thus provide a natural way for studying
the theory beyond the limited scope of perturbative expansions. 
Unfortunately they consist of an infinite tower of coupled integral 
equations relating full $n$-point functions to full $(n+1)$-point functions. 
Thus, the integral equation satisfied by one propagator may involve another
propagator and a three-point vertex. The equation for this vertex may in 
turn include another propagator and a four-point vertex or scattering kernel, 
and so on. Some physically motivated truncation scheme becomes mandatory 
before the infinite tower can be brought to a manageable size. As it has 
been stressed in~\cite{bakerball} the Ward identities of gauge theories 
significantly ease this truncation, since they imply that two-point functions 
(propagators) uniquely determine the longitudinal part of three-point 
functions (vertices).

The \sde\ for the fermion and gauge boson propagators in QCD and QED have 
been studied using different approximations and models. For an excellent 
review see~\cite{review} and references therein. A recurring topic in 
this area is the question of the analytic structure of the propagators. 
For the electron propagator one expects a singularity at $p^2=m_{phys}^2$. 
For a confined particle like the quark it is not so clear what one 
should expect, it depends on what the confining mechanism is. Coleman
~\cite[pp. 378-386]{coleman} has shown that it is perfectly possible for a confined 
particle to have a pole in its propagator at a positive real 
(i.e., physical and time-like) value of $p^2$.

Fermion propagators with branch points at complex conjugate locations 
off the real axis of the variable $p^2$, have been obtained since 
more than twenty years ago for QED~\cite{blatt}, and more recently 
for QCD as well~\cite{review}. This has been studied in great 
detail by Maris~\cite{maristh}. In the case of QED, where we know 
that there must be a singularity at $p^2=m_{phys}^2$, this is 
thought to be an artefact of the approximations with no physical 
significance. For QCD, the absence of singularities on the real 
axis has been thought to be related to quark confinement. 
\c{S}avkli and Tabakin~\cite{savkli} have shown that the imaginary 
part of the value of $p^2$ where the singularity occurs may be 
interpreted as a decay width for a single quark state, and can be related 
to the hadronization distance. However, the fact that in two 
physical situations as different as those of QED and QCD the 
singularities seem to have the same origin~\cite{maristh} makes 
any physical interpretation difficult. The location of these singularities 
poses yet another difficulty: it invalidates the Wick rotation, 
and thus makes the theory as defined in Euclidean metric not equivalent 
to that defined in Minkowski metric. 

In this paper we study the \sde\ for the quark propagator starting from
its Minkowski-space formulation. We do make assumptions about the 
location of the singularities similar to those necessary to justify 
a Wick rotation. However, we arrive at a formulation that allows us 
to solve for the quark propagator in the space-like region (which 
would correspond to a solution in Euclidean space) or, in principle, 
to extend into the time-like region as far as needed.

Our derivation of the \sde\ in a light-cone representation is discussed 
in detail in section \ref{lcsection}. The method is an extension of the 
method used in perturbation theory~\cite{chang&ma} to obtain an infinite 
momentum frame formulation, which in many respects is equivalent to 
the light-cone field theory formulation of perturbative diagrams. 
It is also analogous to the method for obtaining a light-cone 
representation of the Bethe-Salpeter equation from the standard four-
dimensional instant-form, however, as explained in section \ref{lcsection} 
the simple prescription for projecting onto the light cone cannot 
in general be used for the SD equation.

In section \ref{models} we discuss our models.
In sections \ref{integrations} and \ref{tech} we discuss some 
technical aspects of solving the equations and give sample 
results in Minkowski space. In section \ref{results} we discuss 
the results of our model calculations and in section \ref{concl} 
give our conclusions.

\section{ The quark SD Equation and the light-cone.}\label{lcsection}

\hspace{3mm}
In this section we discuss our light-cone formulation the \sde. Since
the light-cone representation of the Bethe-Salpeter (BS) equation for bound
systems is described extensively in the literature, we also briefly review
the BS equation to motivate the present work and explain its applicability.

\begin{figure}
\begin{center}
\epsfig{file=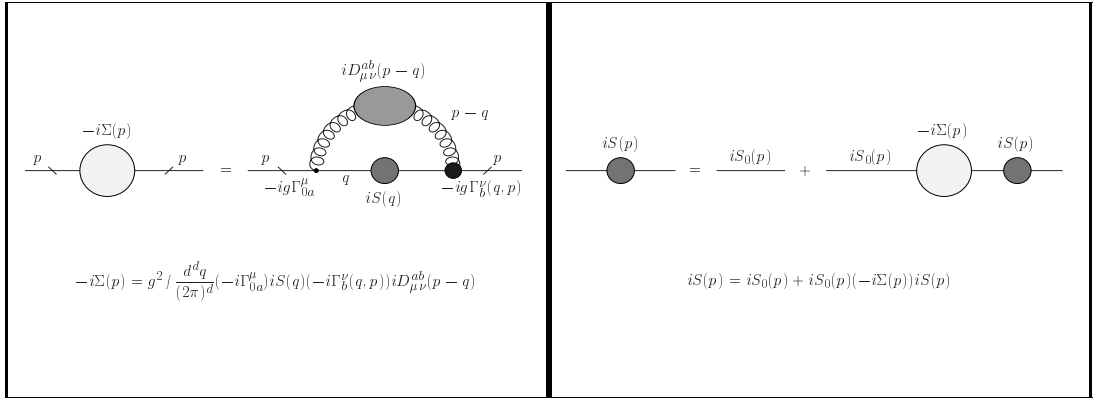,width=412pt,height=150pt}
\caption{Diagrammatic representation of the \sde\ for the quark propagator. 
\label{diagrams}}
\end{center}
\end{figure}

The full dressed quark propagator satisfies the Schwinger-Dyson (SD) equation 
(see Fig~\ref{diagrams}):
\begin{eqnarray}
 S^{-1}(p) & = & 
S_0^{-1}(p)-\raisebox{-0.5ex}{\mbox{\Large $\Sigma$}}(p) \nonumber \\
           &   &                                                 \label{original} \\
\raisebox{-0.5ex}{\mbox{\Large $\Sigma$}}(p)  & = & 
ig^2\int \frac{d^dq}{(2\pi)^d}
\Gamma^{\mu}_{0a}S(q)\Gamma^{\nu}_{b}(q,p)D^{ab}_{\mu \nu}(p-q), \nonumber
\end{eqnarray}
where $\Gamma^{\nu}_{b}(q,p)$ and $D^{ab}_{\mu \nu}(p-q)$ are the dressed
quark-gluon vertex and dressed gluon propagator. Greek letters 
represent Lorentz indices and Latin letters 
stand for color indices. 
The color structure of all quantities in (\ref{original}) is known:
\begin{description}

\item The bare vertex: $\Gamma^{\mu}_{0a}=\gamma^{\mu} \lambda_a/2$

\item The dressed vertex:  $\Gamma^{\nu}_{b}(q,p)=\Gamma^{\nu}(q,p) 
\lambda_b/2$

\item The gluon propagator: $ D^{ab}_{\mu \nu}(k)=\delta^{ab}D_{\mu \nu}(k)$

\end{description}
\noindent so that the color indices can be contracted: $\lambda_a \lambda_b 
\delta^{ab}=16.$ The inverse bare propagator is
 $S_0^{-1}(p)=\dirac{p}-m_c$ , with the 
current quark mass $m_c$ zero in the chiral limit or estimated from PCAC.

The solutions of the \sde\ are of the form
\begin{equation}
\label{sd}
 S^{-1}(p)= A(p^2)\dirac{p} - B(p^2).
\end{equation}

Physically, the most significant aspect of the solution for the two 
functions $A(p^2)$ and $B(p^2)$ is the ratio $M(p^2) = B(p^2)/A(p^2)$,
which is interpreted as the effective mass of the dressed quark propagator.
Since isolated quarks are confined, the interpretation of this mass is
not straight-forward as for the electron SD equation.

The light-cone formulation of quantum mechanics\cite{dirac} starts with
the demonstration that in the light-cone representation one obtains
light-cone Poincare generators, with an interaction-free Lorentz boost in
one direction. One can also show that the analogous result is true in a
light-cone field theory. This enables one to study high momentum transfer
processes, such as form factors in the region above 1 GeV, which is very
difficult in the instant form. Although there was work on the SD equation
three decades ago for perturbative QED~\cite{chang&ma}, for nonperturbative
QCD the problem is quite different, as we show below.
Before discussing the light-cone representation of the \sde\ we review
the well-known light-cone representation of the Bethe-Salpeter 
equation. Since in physical application of the BS amplitudes to form 
factors and transition amplitudes one needs the dressed quark propagator, 
the main motivation of our present work is to formulate a light-cone 
Schwinger-Dyson (LCSD) equation to be used with the light-cone 
Bethe-Salpeter (LCBS) formalism.

There has been a great deal of work on bound state problems in light-cone
representations based on the Bethe-Salpeter (BS) equation. 
For example, the BS equation for the pion in relative coordinates 
has the form
\begin{equation}\label{bs}
\Psi(k) = \int d^4l{\cal K}(k,l)\Psi(l), 
\end{equation}
where ${\cal K}$ is the kernel and $\Psi$ is the BS amplitude. One can obtain
a light-cone representation~\cite{LB} by inserting a delta function to 
include the light-cone on-shell condition and eliminate the $l^-$ variable
\begin{eqnarray}\label{LCBS}
\Psi(x,{\bf k}_{\perp})&=&\int [dy][d^2{\bf l}_\perp]
 {\cal K}(x,{\bf k}_\perp;y,{\bf l}_\perp)\Psi(y,{\bf l}_\perp).
\end{eqnarray}
BS amplitudes from this type of equation have been used to study the
transition from nonperturbative to perturbative regions~\cite{JK,KW}.
We emphasize that for such calculations both on-shell and off-shell
aspects of the SD functions  $A(p^2), B(p^2)$ are needed.
Therefore the prescription of projecting onto the light cone as for 
the BS equation is not appropriate.. We now discuss our approach to the 
the SD equation for the quark propagator on the light cone.

We solve equation (\ref{original}) in a light-cone representation by
using a method introduced originally for perturbation theory. 
In Ref ~\cite{chang&ma} 
S.Chang and S. Ma showed how the rules of light-cone perturbation theory
(LCPT)~\footnote{More accurately, their work refers to the Feynman rules
in the infinite momentum frame, but the difference is of no bearing in 
our present discussion.} can be derived from the usual covariant Feynman 
rules simply by changing into light-cone variables:
\[ q^{\pm}=q^0 \pm q^{3}, \;\; {\bf q}_{\perp}=
\left( q^1,q^2 \right) \]
In simple calculations, like a one-loop self-energy diagram, an important 
feature of the new rules becomes apparent: the range of integration over the
$q^+$ variable becomes finite 
 \[ \int_{-\infty}^{+\infty}dq^+ \longrightarrow \int_0^{p^+}dq^+, \]
where $p^+$ is the ``plus'' (longitudinal) component of the external momentum.
As shown in that paper, this feature is related to the fact that in LCPT 
there are no diagrams with lines going backwards in time and no vacuum 
diagrams (with the exception of zero modes and instantaneous terms in 
fermion propagators). This is a crucial property of light-cone field
theory, since it simplifies tremendously the structure of the vacuum. 

In solving the \sde\ we are faced with the non-perturbative self-energy
diagram seen in Fig.~\ref{diagrams}: 
\begin{equation}\label{genericintegral}
\int \frac{d^dq}{(2\pi)^d} q^{\mu_1}q^{\mu_2}\ldots 
q^{\mu_n}f_Q\left(q^2+i\epsilon\right)f_G\left((p-q)^2+i\epsilon \right)
\end{equation}

Consider first the scalar case $n=0$. Using the variables:
\[ \alpha=q^+/p^+, \;\; s'=q \cdot q, \;\; s=p \cdot p, \;\; 
{\bf q'}_{\perp}={\bf q}_{\perp}-\alpha \; {\bf p}_{\perp}, \] 
the integral in (\ref{genericintegral}) becomes:
\begin{equation}\label{inmyvariables}
\int \frac{ds'd\alpha d^{d-2}{\bf q'}_{\perp}}{2 | \alpha |(2 \pi)^d}
f_Q\left(s'+i\epsilon\right) 
f_G\left(-\alpha^{-1} {\mathcal P}(q^{'2}_{\perp},s,s',\alpha,\epsilon)\right),
\end{equation}
where all the integrals are from minus infinity to plus infinity, and
\[ {\mathcal P}(q^{'2}_{\perp},s,s',\alpha,\epsilon) \equiv
q^{'2}_{\perp}+\alpha (1-\alpha)(-s)+(1-\alpha) s'-i \alpha \epsilon \]
is reminiscent of the similar quantity that appears after the well known 
Feynman trick of combining denominators. Particularly, if 
$s'=m^2_0-i \epsilon$ and $f_G(s)=1/(s-m^2_1)^{a_1}$, then after the 
${\bf q'}_{\perp}-$integration
\[ \alpha^{-1} f_G\left(-\alpha^{-1} {\mathcal P}\right) \rightarrow 
\frac{(-1)^{-a_1} \alpha^{a_1-1}}{\left( 
\alpha (1-\alpha)(-s)+(1-\alpha)m^2_0+\alpha m^2_1-i \epsilon\right)^{a_1}},\]
a very familiar expression.
 
Returning to (\ref{inmyvariables}), let now $s'_Q$ and $s'_G$ 
be singular points of $f_Q$ and $f_G$, respectively. 
Then the integrand in (\ref{inmyvariables}) has singularities at values 
$s'=s'_1$ and $s'=s'_2$ with imaginary parts given by:
\[ Im \left( s'_1 \right) = Im \left( s'_Q \right) - \epsilon, \mbox{ and } 
Im \left( s'_2 \right) = \frac{ \alpha }{\alpha - 1} 
\left[ Im \left( s'_G \right) - \epsilon \right]. \]
It is then clear that if both $s'_Q$ and $s'_G$ are on or below the 
real axis, then the corresponding singularities of the integrand 
will be on opposite sides of the real axis iff 
$ \alpha /\left( \alpha - 1 \right) < 0$, i.e., 
$0< \alpha < 1$. For values of $\alpha$ outside this interval, 
both singularities fall on the same side of the real axis, 
and the contour of the $s'$ integration can be closed with a 
semicircle that doesn't contain either of them. Thus {\em if all the
singularities of the functions $f_Q$ and $f_G$ occur on or below the
real axis, only the interval $(0,1)$ in the integration over $\alpha$
contributes to the integral~\footnote{We are tacitly assuming that the 
integrals are ultravioletly convergent and thus closing the contour with a 
semicircle at infinity introduces no additional contribution.}}. This 
reasoning is only a slight modification of that presented in~\cite{chang&ma}. 
The integral in (\ref{inmyvariables}) then becomes
\begin{equation}\label{lcform}
\frac{1}{4\pi}
\int_{-\infty}^{+\infty}\frac{ds'}{2\pi} \!
\int_0^1 \! d \alpha \!
\int \frac{d^{d-2}{\bf q'}_{\perp}}{(2 \pi)^{d-2}}
\alpha^{-1}f_Q\left(s'+i\epsilon\right) 
f_G\left(-\alpha^{-1}  {\mathcal P}(q^{'2}_{\perp},s,s',\alpha,\epsilon)
\right).
\end{equation}

In dealing with fermion propagators we encounter integrals with 
powers of the momenta in the numerator as shown in (\ref{genericintegral}). 
As discussed in~\cite{chang&ma}, in such cases the previous reasoning 
fails for some components of the integrals (usually referred to as 
``bad'' components). The issue then arises as to whether or not one 
can avoid all such components. In all such cases we are able to avoid 
them and compute only the ``good'' components. The ``bad'' components
 are recovered by the requirements of Lorentz symmetry.

We see that to preserve the rules of light-cone theory we must make 
assumptions about the location of the singularities, much like those 
necessary to justify a Wick rotation. 

A comment on (\ref{lcform}) as compared to a typical calculation of 
a perturbative diagram in a light-cone approach is in order. In most 
light-cone calculations, the integral over $q^-$ (equivalent to the 
integral over $s'$ in (\ref{lcform})) is absent. An on-shell condition 
that determines the value of $q^-$ is used, instead of integrating over 
all values. In other words, the calculations are performed in time-ordered 
perturbation theory (TOPT). This has become so customary that the rules of 
light-cone TOPT are often identified with the rules of light-cone theory. 
The equivalence between TOPT (either light-cone or equal-time) and the 
Feynman rules is usually shown by closing a contour of integration 
to pick up a singularity in a propagator and thus put a particle on 
shell. This can be done in the equal-time formulation (by doing first 
the integral over $q^0$, for example) or in light-cone formulation 
(by doing first the integral over $q^-$). One should refrain from doing 
so in dealing with the \sde. Most certainly, the singularities of 
the Green's functions involved are more complicated than just a pole, 
for one thing: this is not perturbation theory. While such practice comes 
in naturally in the equal-time formulation, where TOPT is not very 
customary, it raises a flag in a light-cone approach. Hopefully, this 
comment addresses that flag.
  
\section{Models}\label{models}

\hspace{3mm}
There has been an extensive program of research on the \sde\ during the
past decade. See \cite{review,review1} for reviews. The usual approach
is to model the gluon propagator and to use symmetries to express the
dressed vertex in terms of the propagators, or to use the free
vertex. The model is constrained by the vacuum condensates. 
We follow this general procedure in our light-cone approach. 
Since our approach allows one to define meson form factors 
for all momentum transfers, which is
not true of instant-form approaches, applications of our solutions to
hadronic properties will be most interesting.

The Lorentz tensor structure of the gluon propagator has the general
form:
\begin{equation}\label{gluongprop}
D_{\mu \nu}(k) = \frac{-1}{k^2+i\epsilon}\left\{
\left( g_{\mu \nu}- \frac{ k_{\mu}k_{\nu} }{k^2+i\epsilon}\right)D(k^2)
+\xi \frac{ k_{\mu}k_{\nu} }{k^2+i\epsilon}\right\}
\end{equation}
here $\xi$ is the gauge parameter. The choice of a model is to pick a value
of $\xi$, such as $\xi = 0 $ for the Landau gauge and the form of the
function D($k^2$). The two essential nonperturbative features that the
gluon propagator must be consistent with are confinement and the vacuum
condensates. It has long been known that with the gluon propagator having
a $1/k^4$ behavior there is confinement\cite{tHooft} and one can fit the 
string tension. On the other hand, it is also known that with the instanton
liquid model \cite{shu} one can fit the quark and gluon condensates. The
two models that we investigate in the present work are the polynomial
model and the instanton model, motivated by these two features of
nonperturbative QCD.

\subsection{Polynomial model}

\hspace{3mm}
We work in Landau gauge with $\xi=0$, and take as our model gluon propagator
\begin{equation}\label{gluongeneric}
D_{\mu \nu}(k) = \frac{-1}{k^2+i\epsilon}
\left( g_{\mu \nu}- \chi \frac{ k_{\mu}k_{\nu} }{k^2+i\epsilon}\right)D(k^2),
\end{equation}
with the parameter $\chi$ introduced to allow us to use a Feynman-like gauge 
(the gauge where $D_{\mu \nu}(k) \propto g_{\mu \nu}$). Thus $\chi=0$ for 
Feynman-like gauge and $\chi=1$ for Landau gauge. See Refs\cite{review,
review1} for discussions the choice of gauge and gauge invariance for \sde\ 
in QCD.

It has been shown elsewhere~\cite{bg} that renormalization group arguments 
yield an approximate relation between the renormalized coupling constant, 
the renormalized gluon propagator and the effective coupling:
\begin{equation} \label{alphaD}
g^2_R D_R(k^2) \approx g_{eff}^2(k^2) = 4 \pi \alpha_{eff}(k^2),
\end{equation}
where the subscript $R$ denotes renormalized quantities, 
and $g_{eff}$ is the effective running coupling constant. 
The renormalized coupling is related to the running coupling by 
\( g^2_R \equiv \left. g^2_{eff}(Q^2)\right|_{Q^2=\mu^2} \), where 
$\mu^2$ is the renormalization point.
 
As discussed in ~\cite[section 6.1]{review}, Eq.(\ref{alphaD}), and 
the fact that in Landau gauge only the combination 
$g^2 D\left(k^2\right)$ enters the \sde, reduces the problem 
of modeling the non-perturbative part of the gluon propagator to that
of modeling the non-perturbative part of $\alpha_{eff}$. We model 
$\alpha_{eff}$ with the expression:
 \begin{equation}\label{alpharunning}
\alpha_{eff}(k^2)= \sum_{l=1}^{N} (-1)^{c_l} \lambda_l 
\left( \frac{ s_0}{k^2+i\epsilon} \right)^{c_l}.
\end{equation}

\begin{figure}
\begin{center}
\epsfig{file=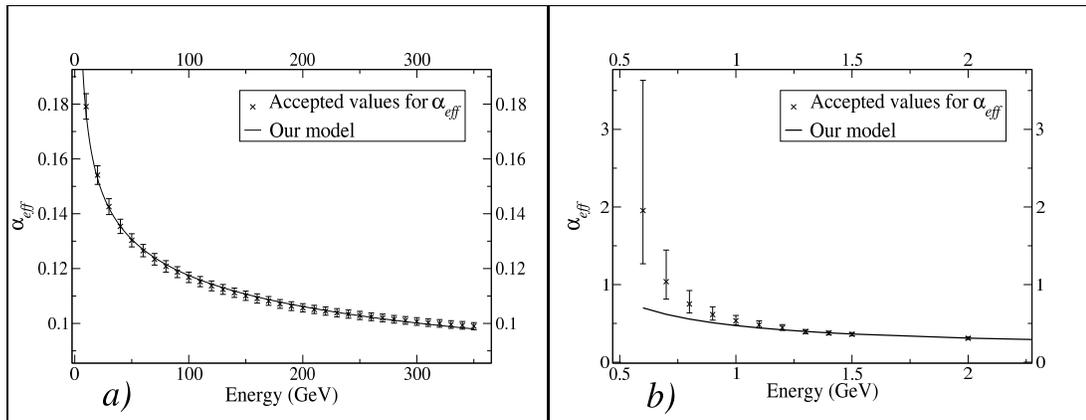,width=412pt,height=159pt}
\caption[Graphs of $\alpha_{\mbox{strong}}$]{Fitting 
$\alpha_{eff}$ with Eq.(\ref{alpharunning}). See text for 
values of parameters. \textit{a)} For energies up to 350 GeV.
\textit{b)} A close-up for lower energies. \label{alphafit}}
\end{center}
\end{figure}

Although the known logarithmic behavior of $\alpha_{eff}$ 
in the high energy limit cannot be fitted accurately with 
this form, a reasonable approximation can be obtained up to 
energies well beyond our main region of interest. Thus, for example, 
with $ N=2,\; s_0=1 \mbox{ GeV}^2, \;\lambda_1= 0.222,\; c_1=0.07,
\; \lambda_2=0.25,\; c_2=0.6 $ the values for $\alpha_{eff}$ 
obtained from our fit are within the accepted error bars as 
published, for example, by PDG\cite{pdg} \footnote{ Numerical data for 
$\alpha_{strong}$ taken from the PDG at http://www-theory.lbl.gov/
\raisebox{-4pt}{\~{}}ianh/alpha/alpha.html} for energies 
from about 1.3 GeV to 350 GeV  [see Fig~\ref{alphafit}]. Fits 
good to higher energies can always be achieved by adding more 
terms in (\ref{alpharunning}) with smaller $c_l$. As can be 
seen in Fig.~\ref{alphafit}, with these parameters, our model 
underestimates the accepted values for $\alpha_{eff}$ in the 
important energy range of a few hundred MeV to about 1.2 GeV, 
although it is widely believed that it overestimates them in 
the deep infrared. 

There is little certainty about what happens to the effective 
coupling or the gluon propagator at very low energies. Some 
results~\cite{brown,bg} suggest that the pole at $k^2=0$ gets 
enhanced and could be as strong as $1/k^4$. This would provide an 
explanation for confinement as shown by 'tHooft~\cite{tHooft}. 
Other results~\cite{alkofer,bowman} suggest that the pole gets 
softer or disappears and that the gluon picks up an effective mass. 
Our model describes an infrared-enhanced gluon propagator, 
although not as strongly as $1/k^4$.

\subsection{Instanton model}

\hspace{3mm}
As we shall discuss in section \ref{results} below, one cannot obtain
the known quark condensates with the polynomial model. On the other hand,
it is known that with a pure instanton model one can obtain the quark
condensate but not the string tension. For this reason we consider a
model with quarks propagating in the instanton medium and in addition the
gluon propagator having a 1/$k^\alpha$ structure to get confining effects
of the far infrared. The starting point is the solution for the instanton 
using the classical action~\cite{bel}, which gives for the instanton color
field
\begin{eqnarray}
\label{instanton}
   A^{inst}_{\mu(x) a} & = & \frac{2 \eta_{a\mu\nu} x_\nu}{x^2 + \rho^2} 
 \\ \nonumber
     G^{inst}(x) \cdot G^{inst}(x) & = & \frac{192 \rho^4}{(x^2 + \rho^2)^4}
\end{eqnarray}
where $\rho$ is the instanton size.  The quark zero modes in the instanton
background~\cite{th}, for the + mode with the instanton at position $z$, are
\begin{eqnarray}\label{zero}
  \Psi_z(x) & = & \frac{\rho}{\sqrt{x^2}\pi (x^2 + \rho^2)^{3/2}}
 \frac{1 + \gamma_5}{2}\gamma_\alpha(x_\alpha-z_\alpha) U,
\end{eqnarray}
where $\rho$ is the instanton size and $U$ is a unitary color-spin matrix. 
From this the widely used model of the quark propagating in
the instanton-antiinstanton medium\cite{pob} was derived.
\begin{eqnarray}\label{pob}
            A_I(p) - 1 & = & 0 \nonumber \\
                 B_I(p) & = & K p^2 f^2(\frac{5}{6} p) \nonumber \\
           f(p) & = & \frac{2}{p} -(3 I_0(p) + I_2(p)) \times K_1(p).
\end{eqnarray}

\begin{figure}
\begin{center}
\epsfig{file=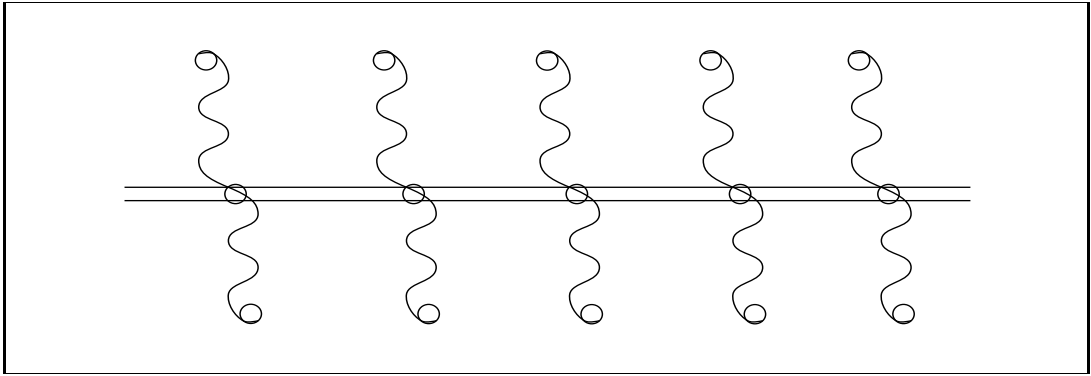,height=141pt,width=412pt}
\caption{Quark propagating in instanton medium \label{quarkinst}}
\end{center}
\end{figure}

  The quark propagating in the instanton-antiinstanton medium is 
illustrated in Fig.~\ref{quarkinst}.

In our model with instantons we include this propagator in the \sde\
by adding it to the free propagator. The justification for this model is
that the instantons give the nonperturbative QCD at the length scale
of about 1/3 fm, while the polynomial model gives the far infrared behavior.
As a simple example of the physics missing if one uses only the polynomial
model for the gluon propagator and ignores instanton effects consider a
delta-function model for the gluon propagator. See, e.g., Ref.\cite{review1}
for the solutions to the \sde\ in such a model. 
The quark and mixed quark condensates are given by
\begin{eqnarray}
\label{condensates}
\lefteqn{\langle0|:\bar{q}(0) q(0):|0\rangle = 
-\frac{3}{4\pi^2} \int_0^{+\infty} dS S \frac{B(S)}{SA^2(S) + B^2(S)} } \nonumber \\
& & \\
\lefteqn{ \langle0|:\bar{q}(0)g \sigma \cdot G(0) q(0):|0\rangle =} \nonumber \\
& & \nonumber \\
& \frac{9}{4 \pi^2} \int_0^{+\infty} dS S \left\{ S
\frac{B(S)(2-A(S))}{SA^2(S) + B^2(S)} +
\frac{81 B(S)[2SA(S)(A(S)-1) + B^2(S)]}{16(SA^2(S) + B^2(S))} \right\}.
 \nonumber
\end{eqnarray}
One can fit the strength of the delta function gluon propagator to obtain
the correct quark condensate. Then from Eqs. (\ref{condensates}) it is
simple to calculate the mixed quark condensate. The result is an order
of magnitude different from the phenomenological value. This is an example
of the need to include both the long-distance and the medium-distance 
nonperturbative QCD effects. We return to this when we discuss the results 
of our calculations. For completeness we include here the equation for $f_{\pi}$
~\cite[Appendix C]{review}
\[ f^2_{\pi}=\frac{3}{4\pi ^2}\int^{+\infty}_0dS \frac{S M(S)}{A(S) [S+M^2(S)]^2}
\left[M(S)-\frac{S}{2}\frac{dM}{dS}\right]. \]
Notice that in these equations $S$ denotes Euclidean momentum.

\subsection{The vertex}

\hspace{3mm}
The results reported in this paper were all obtained with the 
approximation $\Gamma_{\mu}(p,q)=\gamma_{\mu}$ (rainbow 
approximation). The popularity of this approximation is justified 
by its simplicity. It suffers the serious drawback of violating 
the Ward-Takahashi identity (WTI):
\begin{equation}\label{WTI}
(p-q)^{\mu} \Gamma_{\mu}(p,q) = S^{-1}(p)-S^{-1}(q)
\end{equation}
This identity holds in QED, and the forms for the vertex discussed 
below were proposed for QED. Such forms are, however, often used for QCD 
as well, since the analogous identity for QCD (known as the Slavnov-Taylor 
identity) reduces to (\ref{WTI}) if ghost effects are ignored. Ignoring 
these effects is believed to be important in the infrared region.

This identity can be used to express the longitudinal part of 
the vertex in terms of the propagator. Ball and Chiu~\cite{ballchiu} 
have proposed the form:
\begin{equation}\label{generalvertex}
\Gamma_{\mu}(p,q)= \Gamma_{\mu}^{BC}(p,q)+\sum_{i-1}^8
f^i\left(p^2,q^2,p \cdot q\right) T^i_{\mu}(p,q),
\end{equation}
where the $T_{\mu}^i$'s are eight transverse tensors 
(i.e. tensors that satisfy $(p-q)^{\mu} T^i_{\mu}(p,q)=0$, 
and thus don't contribute to the WTI). The form of these 
tensors is given in Eq.(3.4) of~\cite{ballchiu}. The eight 
scalar functions $f^i$ are not constrained by the WTI. 
$\Gamma^{BC}_{\mu}$ is given by:
\begin{eqnarray}\label{BCvertex}
\lefteqn{\Gamma^{BC}_{\mu}(q,p)= \gamma_{\mu} 
\left( A(q^2)(1-\beta_1)+ A(p^2)\beta_1 \right)+} & \nonumber \\
& &  \\
& \hspace{-15pt}
{\dis +(q+p)_{\mu} \left( \dirac{q} \beta_1 + \dirac{p} (1-\beta_1 ) \right) 
\frac{A(q^2)-A(p^2)}{q^2-p^2}-(q+p)_{\mu} \frac{B(q^2)-B(p^2)}{q^2-p^2}}.&
\nonumber 
\end{eqnarray}
This is really a small modification of the vertex proposed by 
Ball and Chiu. In their form $\beta_1=1/2$, but other values are 
often used to explore the effects of the transverse part. Besides 
compromising the symmetry of $\Gamma_{\mu}$, changing the value of 
$\beta_1$ can have serious effects on renormalizability, namely 
because of its effect on the transverse part, as discussed below. 
The WTI itself is satisfied for any value of $\beta_1$, and for any choice 
of $f^i$'s.  The converse is also true: any vertex that satisfies 
the WTI can be expressed through (\ref{generalvertex}, \ref{BCvertex}).
 
The vertex has been further constrained by other studies, in
particular, Burdens and Roberts~\cite{b&r93} list a number of 
requirements $\Gamma_{\mu}$ should satisfy, besides the WTI. See also
~\cite[section 3.7]{review}. Exploiting these requirements, and 
particularly the need to maintain multiplicative renormalizability 
in QED, Curtis and Pennington~\cite{cp} narrowed down the 
last term in Eq.(\ref{generalvertex}) to:
\begin{equation}\label{CPvertex}
\Gamma^{CP}_{\mu}(q,p)=\beta_2 
\left( \gamma_{\mu}(q^2-p^2)-(q+p)_{\mu}(\dirac{q}-\dirac{p}) \right) 
\frac{(q^2+p^2) \left( A(q^2)-A(p^2) \right) }
{(q^2-p^2)^2+\left( M^2(q^2)+M^2(p^2) \right)^2}.
\end{equation}
Here too $\beta_2=1/2$ in the original formulation. Applying the approach 
explained in section \ref{lcsection}, we have found that when using 
\[ \Gamma_{\mu}(q,p) = \Gamma^{BC}_{\mu}(q,p) + \Gamma^{CP}_{\mu}(q,p), \] 
one must set consistently $\beta_1=\beta_2$. This ensures, for example 
that the function $A$ remains free of divergencies in Landau gauge, 
a well established result. Notice that this implies that 
using $\Gamma^{BC}_{\mu}$ by itself is, in general, inconsistent. 

We are currently investigating using this form in our model. Although 
that entails neglecting the effect of ghost fields in the STI, it is 
a significant improvement over the rainbow approximation.

\section{Integrating the equations.}\label{integrations}

\hspace{3mm}
In order to solve (\ref{original}) in our model, we need to calculate 
integrals of the type (here $a_l=1+c_l$):
\begin{eqnarray}\label{Cintegrals}
\lefteqn{ \dis \lim_{ \epsilon \to 0^+ } \int \frac{d^{d}q}{(2\pi)^d} 
\frac{ f_Q(q^2+i \epsilon) }
{ \left((p-q)^2-m^2_1+i\epsilon \right)^{a_l}} 
\left\{ 1,\: q^{\mu},\: q^{\mu} \, q^{\nu} \right\}
\equiv }\nonumber \\
& & \\
& \mbox{\hspace{-30pt}} \left\{C_{00}(p^2;f_Q,a_l), 
p^{\mu} C_{10}(p^2;f_Q,a_l),
p^{\mu} p^{\nu} C_{20}(p^2;f_Q,a_l) + 
g^{\mu \nu} C_{01}(p^2;f_Q,a_l)\right\},
& \nonumber
\end{eqnarray}
where we have exploited the Lorentz structure to define the scalar
 quantities $C_{rj}$.

These integrals include as a particular case the well known integrals:
\begin{eqnarray}\label{Iintegrals}
\lefteqn{ \dis \lim_{ \epsilon \to 0^+ } \int \frac{d^{d}q}{(2\pi)^d} 
\frac{\left\{ 1,\: q^{\mu},\: q^{\mu} \, q^{\nu} \right\}}
{\left(q^2-m^2_0 +i \epsilon \right)^{a_0} \left((p-q)^2-m^2_1+i\epsilon 
 \right)^{a_l}} \equiv } \nonumber \\
& & \\
& \mbox{\hspace{-25pt}} \left\{I_{00}(p^2;a_0,a_l), 
p^{\mu} I_{10}(p^2;a_0,a_l),
p^{\mu} p^{\nu} I_{20}(p^2;a_0,a_l) + g^{\mu \nu} 
I_{01}(p^2;a_0,a_l)\right\}.
& \nonumber
\end{eqnarray}

For $(r,j)=(0,0),(1,0),(2,0),(0,1)$, the $I's$ are known to be:
\begin{eqnarray}\label{Iresult}
\dis \lefteqn{I_{rj}(s;a_0,a_l)=\frac{i (-1)^{-(a_0 + a_l)}
(-1/2)^j}{(4 \pi )^{d/2}}
\myGamma{a_0 + a_l - d/2 -j}{a_0, a_l} \times } \nonumber \\
& & \\
& \dis \int_0^1 d \alpha \frac{ \alpha^{a_l+r-1} (1-\alpha)^{a_0-1}}
{( \alpha (1-\alpha)(-s)+(1-\alpha)m_0^2 + \alpha m_1^2 -i\epsilon)^{a_0 + 
 a_l - d/2 - j}}, & \nonumber
\end{eqnarray}
where we have used the compact notation
\[ \myGamma{a_1,a_2, \ldots, a_n}{b_1,b_2, \ldots, b_m}=
\frac{\Gamma(a_1)\Gamma(a_2)\ldots\Gamma(a_n)}{\Gamma(b_1)\Gamma(b_2)\ldots\Gamma(b_m)} \;. \]

For the $C's$, we proceed as explained in section~\ref{lcsection} to get:
\begin{eqnarray}\label{Cresult}
\dis \lefteqn{\mbox{\hspace{-0.5in}} C_{rj}(s;f_Q,a_l)=
\frac{i (-1)^{- a_l}(-1/2)^j}
{(4 \pi )^{d/2}}
\myGamma{a_l + 1 - d/2 -j}{a_l} 
\int_{-\infty}^{+\infty} \frac{f_Q(s'+i\epsilon)ds'}{2 \pi i} } \nonumber \\
& & \\
& \dis \int_0^1 d \alpha \frac{ \alpha^{a_l+r-1}}
{( \alpha (1-\alpha)(-s)+(1-\alpha)s'+ \alpha ( m_1^2 -i\epsilon) )^
{a_l + 1 - d/2 - j}}.  & \nonumber
\end{eqnarray}

At this point we must make sure that if we set 
$f_Q \left( s' \right) = 1 / \left(s'-m^2_0 \right)^{a_0}$, then (\ref{Cresult}) 
agrees with (\ref{Iresult}). This is most easily done by closing the contour 
of integration of the $s'$ variable with an infinite semicircle in the {\em lower} 
half of the complex plane to pick up the singularity~\footnote{If $a_0$ is not an 
integer, then this singularity will be a branch point 
with its corresponding branch cut, as opposed to a pole.} in $f_Q$. 
This procedure, however, uses the analytic form of $f_Q$. 
We need a procedure that would rely only on the numerical values of $f_Q$. 
We perform the integrations by closing our contour with an 
infinite semicircle on the {\em upper} half of the complex plane 
to pick up the singularity (in $s'$) of the expression on the second line 
of (\ref{Cresult}). For $m_1=0$, the $I's$ reduce to~\footnote{The expansion 
of this expression about $d=4$ is probably more familiar to the 
expert in dimensional regularization.}:
\begin{eqnarray}\label{Inomass}
\lefteqn{I_{rj}(s;a_0,a_l)=} \nonumber \\
& & \nonumber \\
& \dis \frac{i (-1)^{-(a_0 + a_l)}(-1/2)^j}{(4 \pi )^{d/2} (m_0^2)^{a_0+a_l-d/2-j}} 
\myGamma{a_0 + a_l - d/2 -j, a_l+r,d/2+j-a_l}{a_0, a_l, d/2+j+r} & \\
& & \nonumber \\ 
& \dis \hyp(a_0+a_l-d/2-j,a_l+r;d/2+j+r;s/m_0^2). & \nonumber
\end{eqnarray}
For the $C's$:
{\small \begin{eqnarray}\label{Cnomass}
\lefteqn{C_{rj}(s;f_Q,a_l)=-\frac{i(-1)^{-a_l}(-1/2)^j}
{(4 \pi )^{d/2}\Gamma[a_l]} \times  } \nonumber \\
& & \nonumber \\
& \hspace{-60pt} \left\{ 
\begin{array}{l@{\hspace{-20mm}}r}
{\dis \myGamma{b}{1+b-a}(-s)^{-b}\int_0^s ds' f_Q(s')(-s')^{b-a}
\hyp(1+b-c,b;1+b-a;s'/s)+} &  \\
&   \\
{\dis + \myGamma{b,c-b}{c,1-a}\int_s^{-\infty} ds' f_Q(s')(-s')^{-a}
\hyp(a,b;c;s/s')}  & s<0  \\  
&     \\
&     \\
{\dis \myGamma{b,c-b}{c,1-a}\int_0^{-\infty} ds' f_Q(s')(-s')^{-a}} &  s=0  \\
&     \\
&     \\
{\dis \myGamma{c-b}{1+c-a-b} s^{1-c}\int_s^0 ds' f_Q(s') 
(s-s')^{c-a-b} (s')^{b-1} \hyp(1-a,1-b;1+c-a-b;1-s/s')+}  &  \\
&  \\
{\dis + \myGamma{b,c-b}{c,1-a}\int_0^{-\infty} ds' f_Q(s')(-s')^{-a}
\hyp(a,b;c;s/s')} & s>0 , 
\end{array} \right.  & 
\end{eqnarray}}
where $a=a_l+1-d/2-j, \; b=a_l+r, \; c=d/2+j+r$.

\section{Sample calculations in Minkowski space.} \label{tech}

\hspace{3mm}
  In this section we discuss sample calculations in Minkowski space.
The ability to work in Minkowski space is one of the important features
of our light-cone formulation of the SD equation. in this section, however,
we only use polynomial models, which allow convenient Minkowski space forms.
As we shall see, the infra-red properties in these models are not consistent
with the known condensates, and solutions with instanton contributions to
the gluon propagator are much more satisfactory. Since the instanton forms
have been obtained in Euclidean space, however, there is no unambiguous
Minkowski space formulation. We discuss our results with models including
instantons in the following section.

  Eq.(\ref{Cnomass}) is a generalization of the 
well known result Eq.(\ref{Inomass}) to a wider class of 
functions $f_Q$. The light-cone approach used, and in particular
the treatment it makes of the singularities, seldom goes without 
some aftertaste. Here is a well known example where using  
light-cone variables leads to a dead end~\footnote{O.L. 
wishes to thank Dr. Matthias Burkardt for bringing
this to his attention.}:
\[ \lim_{\epsilon \to 0^+} \int \frac{dk_0 dk_1}
{(k_0^2-k_1^2-m_0^2+i\epsilon)^n} \;\;\; n > 1 \]
This integral is perfectly convergent and can be computed, for 
example, by closing the contour in the $k_0$-integration with a
 semicircle at infinity. The integral is, of course, non-vanishing. 
Using light-cone variables, one would have the integral:
\[ \lim_{\epsilon \to 0^+} \int \frac{dk^+ dk^-}
{2 (k^+k^- - m_0^2+i\epsilon)^n} \;\;\; n > 1 \]
If we now think of closing the contour of, say the $k^-$-integration, 
we see that the singularity occurs at $(m_0^2-i\epsilon)/k^+$. But then
 the singularity can always be avoided for any $k^+\ne 0$. Only for $k^+=0$
 is the $k^-$-integration non-vanishing, but then, it must have a delta
 function type singularity.

The purpose of this section is test Eq.(\ref{Cnomass}) 
against such anomalies, as well as to point out some important 
technicalities. We input a ``first guess'' for the quark propagator: 
$ f_Q(s')=1/(s'-m_0^2) $ with $m_0=2 \mbox{ GeV}$, and use the model 
for the gluon propagator described in section \ref{models}:
\[ g_R^2 D_R(k^2)= 4 \pi \left\{ 
(-1)^{c_1} \lambda_1 \left( \frac{s_0}{k^2+i\epsilon} \right)^{c_1}+
(-1)^{c_2} \lambda_2 \left( \frac{s_0}{k^2+i\epsilon} \right)^{c_2} \right\}, \]
where $s_0=1 \mbox{ GeV}^2, \; c_1=0.07, \; \lambda_1=0.222, 
\; c_2=0.6, \; \lambda_2=0.25$. We also discuss the ultravioletly 
divergent case $c_1=0$. For the vertex we set 
$\Gamma^{\nu}=\gamma^{\nu}$. We use Eq (\ref{Cnomass}) to compute 
the functions $A \mbox{ and } B$ with this input. 
We exploit the fact that for the first iteration, the integrals 
could also be computed analytically using Eq.(\ref{Inomass}). 
We present graphs comparing the two outputs. We stress that 
these are not solutions to the \sde\ but merely the output 
of running the iterative procedure once. Solutions to the 
\sde\ are presented in section \ref{results}.

\begin{figure}
\begin{center}
\epsfig{file=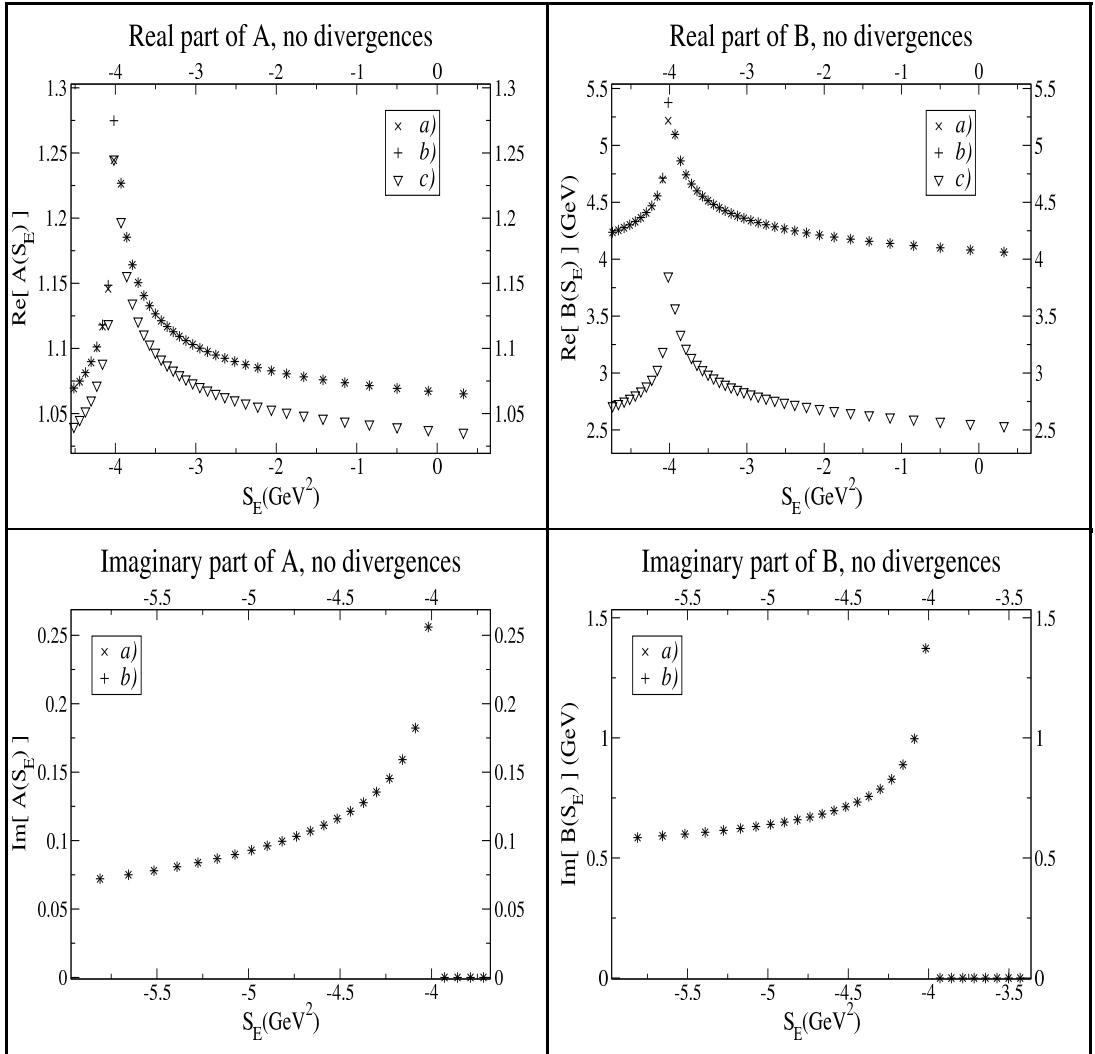,width=412pt,height=400pt}
\caption{The quark propagator after one iteration compared with
 the analytic result. {\em a)} analytic result (\ref{Inomass}); 
numerical integrations of (\ref{Cnomass}):{\em b)} with tail 
contribution, {\em c)} without tail contribution. Notice the 
appearance of a singularity and an imaginary part.
\label{convergent}}
\end{center}
\end{figure}
 
Keeping with tradition, the graphs are shown with a reversed 
x-axis. We use the variable $S_E \equiv -s$, 
equivalent to the Euclidean invariant momentum squared. 
We show, however, negative values of $S_E$, which lay 
outside the usual Euclidean space, and represent the time-like 
region. The numerical integration is carried out up to $S_{E_{MAX}}=64 
\mbox{ GeV}^2$ and an estimate of the contribution of the remaining
 integration out to infinity (``tail contribution'') is added. 
As will be seen on the graphs the consequences of omitting 
this tail contribution are numerically important.

In fact, in order to reproduce the numerical values of 
Eq.(\ref{Inomass}), Eq.(\ref{Cnomass}) demands an elaborate 
integration procedure. On one hand there is the ever present 
fact of having to integrate over all energies. Renormalizability 
tells us that there is much freedom in dealing with the very high 
energy region (it can even be dropped, as in cut-off 
renormalization schemes), and whatever numerical effect that 
has on the integrals can be compensated by readjusting a finite 
number of parameters. From this point of view the effect of omitting 
the tail contribution, although numerically important, should be 
of no physical consequence. Eq.(\ref{Inomass}), however, represents 
the most accepted treatment of the very high energy regime: 
the integrals are carried out to infinity without the introduction 
of cut-offs or form factors that could spoil the symmetries of 
the theory. Divergencies, when they exist, are treated with the 
dimensional regularization method. In order to have Eq.(\ref{Inomass}) 
as a reference, we choose to carry out the integral in Eq.(\ref{Cnomass}) 
out to infinity as well. In a numerical procedure this means adding 
an estimate of the tail contribution. On the other hand, we have a 
 ``floating singularity'' in the integrand (a singularity at  
$S_E=S'_E$, as opposed to one at a fixed point on the grid). 
This is particularly important for the infrared-enhanced terms 
(those with larger exponent $c_l$). Given the strength of the 
singularity at $S_E=S'_E$, the contribution from the region 
$S_E \approx S'_E$ is important. Sampling enough values of $S'_E$ 
immediately to the left of $S_E$, is impossible when 
$S_E \approx S_{E_{MIN}}$, unless the grid is refined each time 
(the analogous problem around $S_{E_{MAX}}$ gets handled by the 
addition of the tail contribution). This can contribute to inaccuracies 
in the very low energy region, a most undesirable effect. Refining 
the grid with each iteration, however, can easily lead to exponential 
growth of the computational time.

\begin{figure}
\begin{center}
\epsfig{file=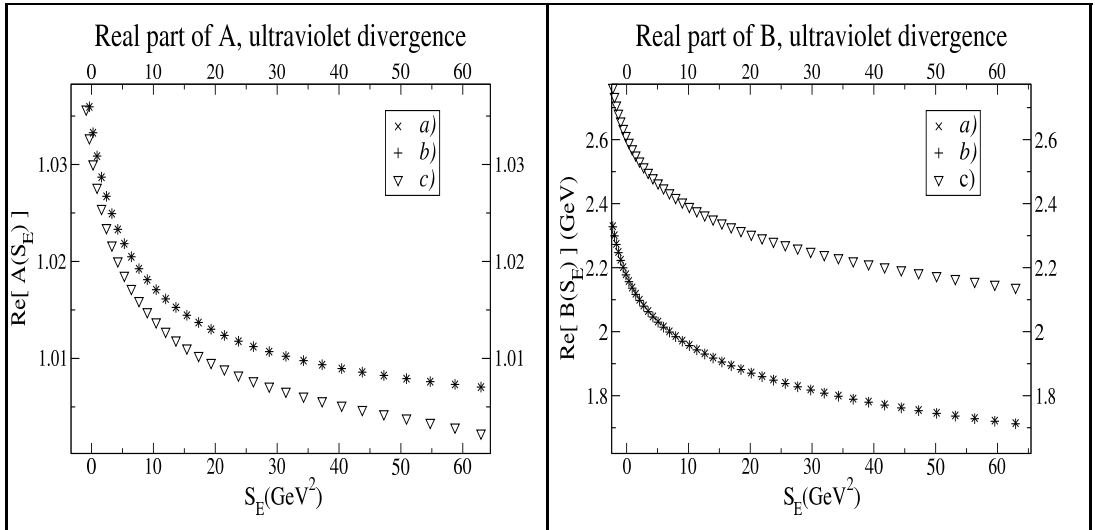,width=412pt,height=200pt}
\caption{The numerically renormalized quark propagator 
after one iteration in an ultraviolet divergent case, 
compared with the analytic \MS\ result. {\em a)} analytic 
result (\ref{Inomass}); numerical integrations of (\ref{Cnomass}):
{\em b)} with tail contribution and \MS\ subtraction, 
{\em c)} without tail contribution or \MS\ subtraction.}
\label{divergent}
\end{center}
\end{figure}

One solution is to use a model propagator with no infrared 
enhancement, and that falls off rapidly at large energies,
 e.g., a Gaussian model. We don't feel, however that Gaussian 
models are appropriate in the time-like region, and one of the 
purposes of this paper is to explore the effects of a tunable 
infrared enhancement. Our solution is to sample numerically only 
the part of the integrand involving the fermion propagator and the vertex, 
and to treat the part coming from the boson propagator 
(and containing the floating singularity) analytically. 
In the language of numerical methods, it is much along the 
lines of Gaussian quadratures. With this, and the addition of
the tail contribution, the procedure becomes accurate, and 
reasonably fast.

The graphs in fig.~\ref{convergent} show the real and imaginary 
parts of the functions $A \mbox{ and } B$ after one iteration 
using the parameters above.  As can be seen from the graphs, 
the procedure is quite accurate (provided the estimate of 
the tail contribution is added!). The relative error is of the 
order of thousandths of a percent, although it is somewhat larger 
right before (but not after!) the singularity at $S_E=-m_0^2$. 
The imaginary parts are obtained by simply applying the correct 
$i\epsilon-$procedure around the singularity in $f_Q$.

Figure~\ref{divergent} shows the {\em renormalized} real 
parts of the same functions, with the crucial difference of setting 
$c_1=0$, rather than $c_1=0.07$. This causes an ultraviolet 
divergence. The imaginary parts in this calculation contain no 
divergence, so they remain basically the same as in the convergent 
case, and thus are not shown. In this case, the analytic results are
renormalized in the \MS\ scheme. The 
numerical calculation was performed with $d=4+2e$, 
where $e=-10^{-7}$. The divergence comes in the tail 
contribution (which is almost constant for the divergent term), 
and it is canceled by the $1/e$ subtraction used in the \MS\ 
scheme. The well known result that $A$ is free of ultraviolet 
divergences in Landau gauge renders this function insensitive 
to whether or not the subtraction is performed. Addition of the tail 
contribution still significantly affects the non-divergent term. 
Even for $B$, it is possible to obtain a finite result by just 
dropping the tail contribution (cut-off renormalization), but we 
choose to integrate to infinity and perform the $1/e$ 
subtraction. We do not include in this paper solutions to the \sde\ 
involving this renormalization procedure, but thought it appropriate 
to show how such solutions could be obtained. We point out that 
dimensional regularization is not believed to give physically 
correct results in non-perturbative calculations. This method 
could be used to explore that question.

\section{Results with instantons contributions}\label{results}

\hspace{3mm}
In the present section we treat models that contain instanton contributions.
Since there are a variety of possible models, we present results 
obtained with different approaches.
On one hand, we want to study Eq.(\ref{original}) in its original form. 
On the other hand we want include the effects of the propagation 
in the instanton-antiinstanton medium, contained in (\ref{pob}). 
We compare three different alternatives for including the effects 
of (\ref{pob}) in (\ref{original}). All in all we study four different
equations: 
\begin{eqnarray}
S^{-1}(p) & = & S^{-1}_0(p)-\raisebox{-0.5ex}{\mbox{\Large $\Sigma_{SD}$}}
\left(S(p)\right) \label{noinst}         \\
\nonumber \\
S^{-1}(p) & = & S^{-1}_I(p)-\raisebox{-0.5ex}{\mbox{\Large $\Sigma_{SD}$}}
\left(S(p)\right) \label{instout}        \\
\nonumber \\
S^{-1}(p) & = & S^{-1}_0(p)-\raisebox{-0.5ex}{\mbox{\Large $\Sigma_{SD}$}}
\left(S(p)+S_I(p)\right) \label{instin}  \\
\nonumber \\
S^{-1}(p) & = & S^{-1}_I(p)-\raisebox{-0.5ex}{\mbox{\Large $\Sigma_{SD}$}}
\left(S(p)+S_I(p)\right), \label{install}
\end{eqnarray}
where $S_0$ is the bare propagator, $S_I$ is the propagator 
given by (\ref{pob}), $S$ is the full dressed propagator, 
and {\Large $\Sigma_{SD}$} is the quark self-energy, thought of here 
as of a function of $S(p)$. The subscript $SD$ 
is introduced here to distinguish this self-energy from the kernel of a 
different equation, discussed below. Thus, Eq.(\ref{noinst}) is just 
Eq.(\ref{original}) and does not explicitly include the instanton effects 
contained in Eqs. (\ref{pob}), while Eqs.(\ref{instout}, \ref{instin}, \ref{install}) 
propose three different ways in which these effects could be included. 
The form of the function $ \raisebox{-0.5ex}{\mbox{\Large $\Sigma_{SD}$}} 
\left( S(p) \right) $ (see Fig.~\ref{diagrams} and Eq.(\ref{original})) 
depends on the model chosen for the gluon propagator, the gauge and the 
vertex. We use two different sets of parameters for the gluon propagator 
and the rainbow approximation for the vertex. All calculations were performed 
in Landau gauge and in the chiral limit ($m_c=0$). The parameters for the gluon 
propagator are:\\

set a: $ \left\{
\begin{array}{ll}
\lambda_1=0.222 & c_1=0.07 \\
\lambda_2=0.25  & c_2=0.6
\end{array} \right.$, and set b: $ \left\{ 
\begin{array}{ll}
\lambda_1=0.222 & c_1=0.07 \\
\lambda_2=1.5   & c_2=0.85
\end{array} \right. $

\begin{figure}
\begin{center}
\epsfig{file=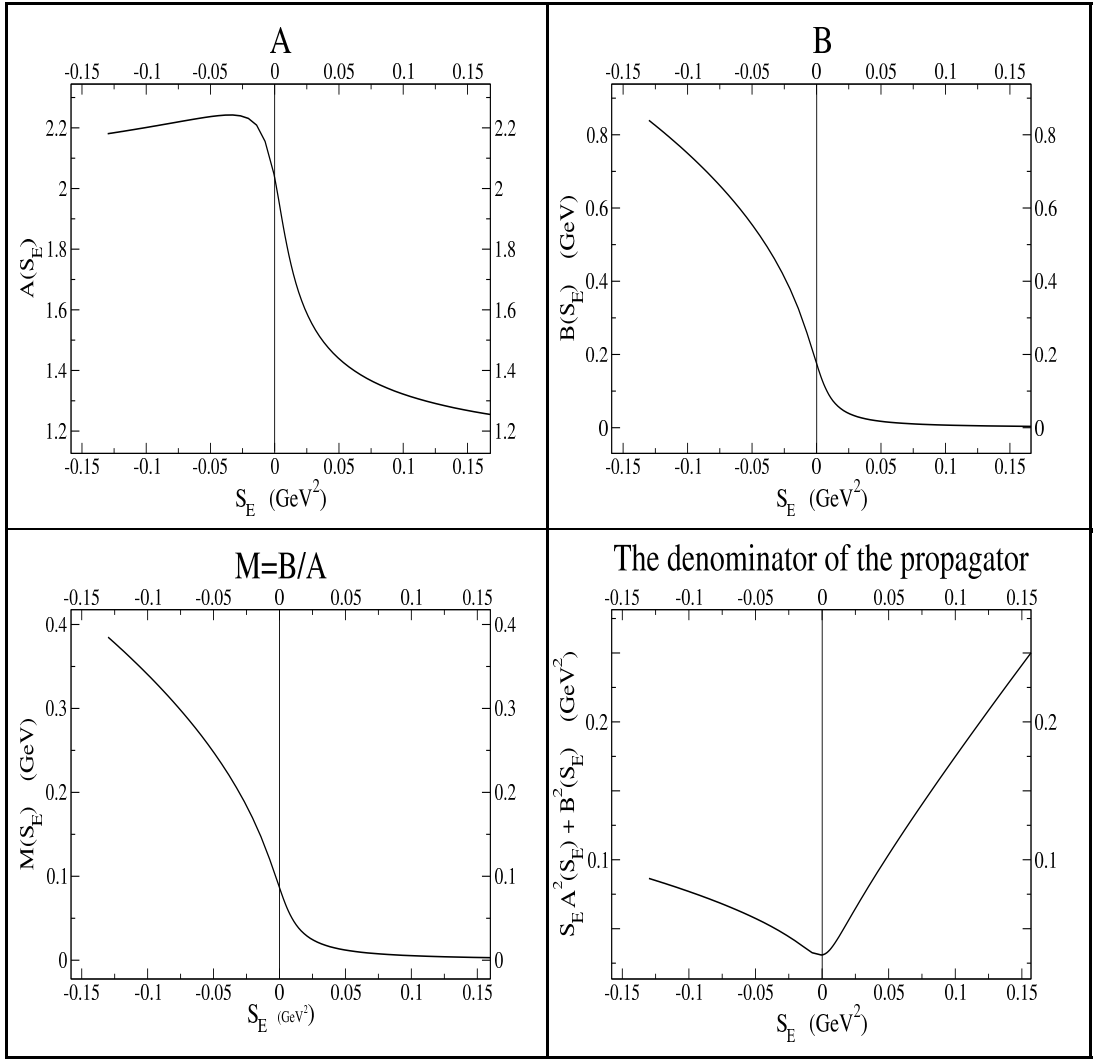,width=412pt,height=400pt}
\caption{Solutions to Eq.(\ref{noinst}) using set a. The left part of
the graphs ($S_E<0$) represents the time-like region.}
\label{TLseta}
\end{center}
\end{figure}

\begin{figure}
\begin{center}
\epsfig{file=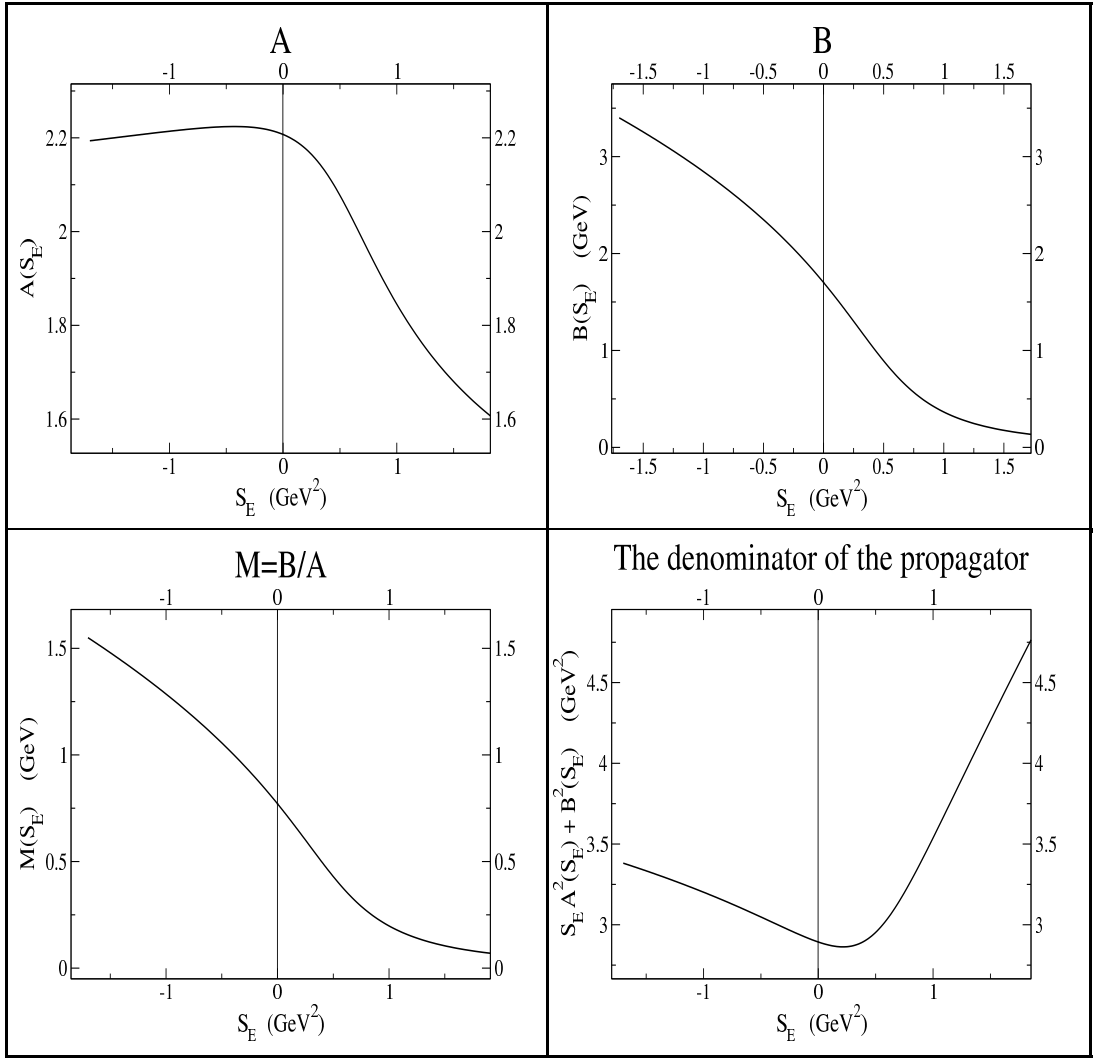,width=412pt,height=400pt}
\caption{Solutions to Eq.(\ref{noinst}) using set b. The left part of
the graphs ($S_E<0$) represents the time-like region.}
\label{TLsetb}
\end{center}
\end{figure}

In the table below we compare the different solutions in terms of the 
values they yield for the quark condensate, the mixed quark condensate, 
$f_{\pi}$, and the value of $M=B/A$ at $p^2=0\;$\footnote{This value is 
often compared to the constituent quark mass ($M_Q$) of the 
non-relativistic quark model (NRQM). As we show in Figs.~\ref{TLseta} 
and \ref{TLsetb}, the solutions to the \sde\ vary quite rapidly upon 
entering the time-like region. The value $M_Q$ should therefore, at 
best, be regarded as a rough guide to the value of $B/A$ at $p^2=0$.
\label{MQ}}.
\begin{center}
\begin{minipage}[t]{372pt}
\hspace{-36pt}
\begin{tabular}{|c||c|c||c|c||c|c||c|c|}
\hline
\multicolumn{9}{|c|}{Vertex: $\Gamma^{\nu}=\gamma^{\nu}$ Rainbow approximation} \\
\hline
model                                                                            & 
\multicolumn{2}{c||}{$(-\langle:\bar{q}q:\rangle)^{1/3}$}                        & 
\multicolumn{2}{c||}{$(-\langle:\bar{q}g \sigma \cdot G q:\rangle)^{1/5}$}       &
\multicolumn{2}{c||}{$f_{\pi}$}                                                  &
\multicolumn{2}{c|}{$M(0)$}                                                      \\
                            & 
\multicolumn{2}{c||}{(MeV)} & 
\multicolumn{2}{c||}{(MeV)} &
\multicolumn{2}{c||}{(MeV)} &
\multicolumn{2}{c|}{(MeV)}  \\
\cline{2-9}
                    & Set a & Set b & Set a  & Set b & Set a & Set b & Set a & Set b  \\
\hline
Eq.(\ref{noinst})   & 58    & 288.5 & 313.5 & 1111.5 & 13    & 89    & 86.5  & 770.5  \\
\hline
Eq.(\ref{instout})  & 221   & 227.5 & 558   & 906.5  & 91.5  & 117.5 & 534   & 1084.5 \\
\hline
Eq.(\ref{instin})   & 171.5 & 153.5 & 509   & 975.5  & 58.5  & 54.5  & 308   & 752.5  \\
\hline
Eq.(\ref{install})  & 219   & 159.5 & 624   & 1001   & 91    & 60    & 607   & 804    \\
\hline \hline
\multicolumn{9}{|c|}{For comparison}\\
\hline
Eq.(\ref{pob})                     &
\multicolumn{2}{c||}{216.7}        &
\multicolumn{2}{c||}{456.5}        & 
\multicolumn{2}{c||}{86}           & 
\multicolumn{2}{c|}{417.6}        \\
\hline
Other                            &
\multicolumn{2}{c||}{200-250}    &
\multicolumn{2}{c||}{400-600}    &
\multicolumn{2}{c||}{92}         &
\multicolumn{2}{c|}{$\sim 300$} \\
\cline{2-9}
calc.                                                   & 
\multicolumn{4}{c||}{Sum rules, lattice QCD}            &  
\multicolumn{2}{c||}{Exper.}                            &   
\multicolumn{2}{c|}{NRQM\footnote{see footnote \ref{MQ} at the 
bottom of page \pageref{MQ}.}}                           \\
\hline 
\end{tabular}
\end{minipage}
\end{center}
Most of the results obtained with Eq.(\ref{noinst}) (no instanton effects) 
set a, appear too low. This is most likely due to the fact that 
the corresponding coupling is too weak in the crucial region of a 
few hundred MeV to about 1.2 GeV. Inclusion of the instanton effects 
seem to improve matters considerably, perhaps with the exception of 
Eq.(\ref{instin}).

A plausible explanation for this maybe as follows: the solution (\ref{pob}) 
was also obtained from a self consistent equation~\cite{pob}, 
which we write symbolically as:
\begin{equation}\label{pobeqn}
S^{-1}_I(p) = S^{-1}_0(p)-\raisebox{-0.5ex}{\mbox{\Large $\Sigma_I$}}
\left(S_I(p)\right).
\end{equation}
One could think of including both types of one-particle-irreducible 
diagrams into the self-energy:
\begin{equation}\label{hypeqn1}
S^{-1}(p) = S^{-1}_0(p)-\raisebox{-0.5ex}{
\mbox{\Large $\left[ \Sigma_{SD}+\Sigma_I \right]$}}
\left(S(p)\right).
\end{equation}
Solving (\ref{pobeqn}) for $S^{-1}_0$ and substituting into (\ref{hypeqn1})
\begin{equation}\label{hypeqn2}
 S^{-1}(p) = S^{-1}_I(p)-\raisebox{-0.5ex}{\mbox{\Large $\Sigma_{SD}$}}
\left(S(p)\right)-\raisebox{-0.5ex}{\mbox{\Large $\Sigma_I$}}
\left(S(p)-S_I(p)\right)
\end{equation}
\begin{figure}
\begin{center}
\epsfig{file=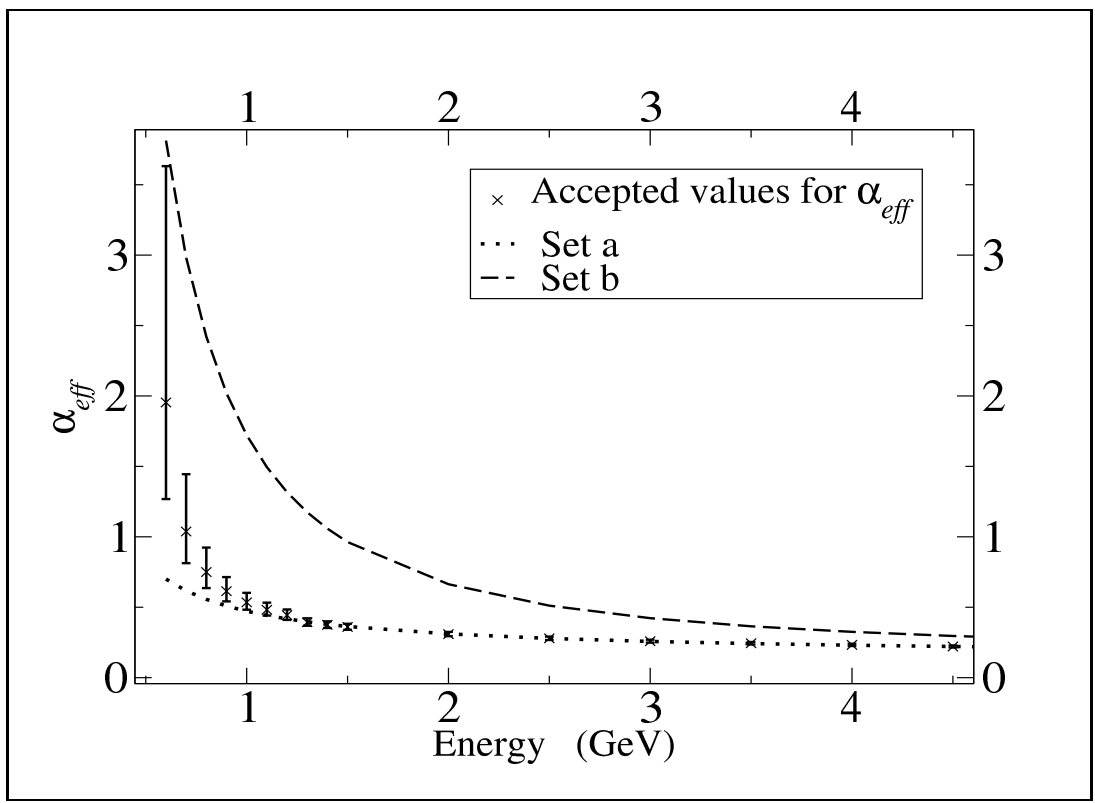,width=412pt,height=300pt}
\caption{Two different fits to $\alpha_{eff}$\label{ab-alpha}}
\end{center}
\end{figure}
Eq.(\ref{hypeqn2}) contains the effects of the quark propagating in the 
instanton medium as suggested in section \ref{models}. An important feature 
is the presence of $S^{-1}_I$ instead of $S^{-1}_0$ in the inhomogeneous term,
which occurs also in Eqs. (\ref{instout}, \ref{install}), 
but not in Eq.(\ref{instin}). In fact, Eq.(\ref{instout}) is an approximation 
to Eq.(\ref{hypeqn2}) up to terms of order 
$\raisebox{-0.5ex}{\mbox{\Large $\Sigma_I$}}\left(S(p)-S_I(p)\right)$, which can be 
considered small, as opposed to $S^{-1}_0-S^{-1}_I$, which obviously is not small.
The results obtained with Eq.(\ref{noinst}) (no instanton effects) set b, appear 
closer to the correct values, often overestimating them. This is most likely due 
to the fact that set b contains stronger infrared enhancement and overestimates the 
coupling the intermediate region. A comparison of both sets in terms of how they fit 
the accepted values of $\alpha_{eff}$ is given in Fig.~\ref{ab-alpha}. Inclusion 
of the instanton effects with set b, in most cases, do not lead to worse 
overestimation, as could be naively expected. With set a we have a situation 
where the polynomial part of our model is contributing very little in the 
intermediate region (which we believe to be crucial to the quantities we 
are computing), and the instanton effects bring in the bulk of the contribution. 
With set b both parts are contributing significantly, but their effects do 
not add. Thinking along the lines of Eq.(\ref{hypeqn1}), this probably means that 
the two types of diagrams are of different nature and not always interfere 
constructively. This suggests that the instanton effects are not being 
double-counted with this procedure, although such conjecture needs further testing.

Figures~\ref{TLseta} and \ref{TLsetb} show graphs of the solutions 
obtained using set a and set b, respectively, and Eq. (\ref{noinst}). 
The focus is on the low energy space-like region (right half of the graphs)
and the cross-over to the time-like region (left half of the graphs). Already 
at small time-like energies (the graphs show energies up to about 
360 MeV for set a and 1.3 GeV for set b) the behavior of the functions 
has changed dramatically from their behavior in the space-like region. 
The function $A$, for example, stops growing; and, say, the function $M$ 
for set a, (see Fig~\ref{TLseta}) which has barely risen from its asymptotic 
value of $M(p^2) \to m_c=0$ at large space-like $p^2$ to about 90 MeV at 
$p^2=0$, is already close to 400 MeV at $p^2 \approx (360 MeV)^2$.

\begin{figure}
\begin{center}
\epsfig{file=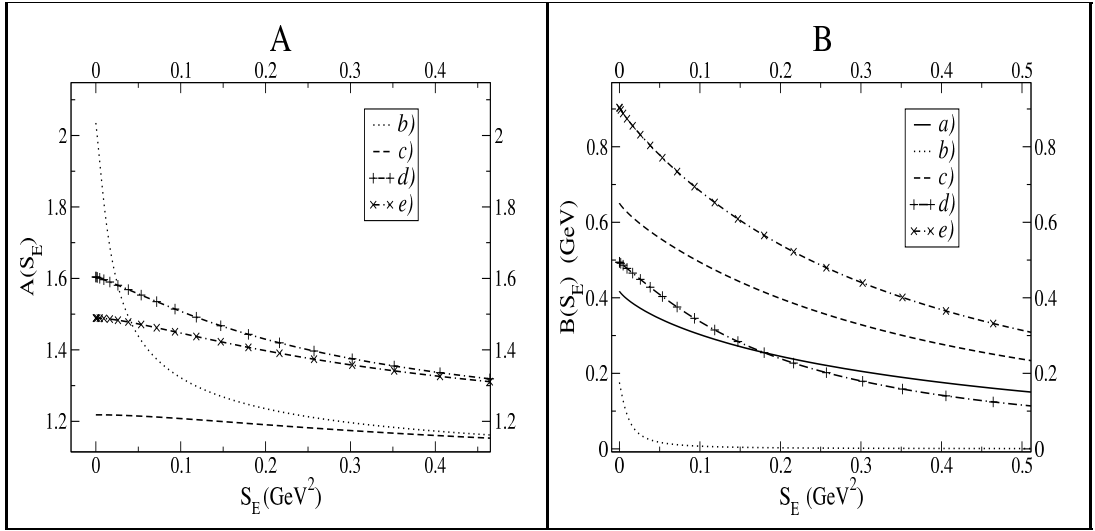,width=412pt,height=200pt}
\caption{Solutions to the \sde\ using set a in the space-like
region. {\em a)} Result Eq.(14) from Ref[26], {\em b)}
Eq.(26), {\em c)} Eq.(27), {\em d)} Eq.(28),
(e) Eq.(29). In (a) $A=1$.} 
\label{SLseta}
\end{center}
\end{figure}

\begin{figure}
\begin{center}
\epsfig{file=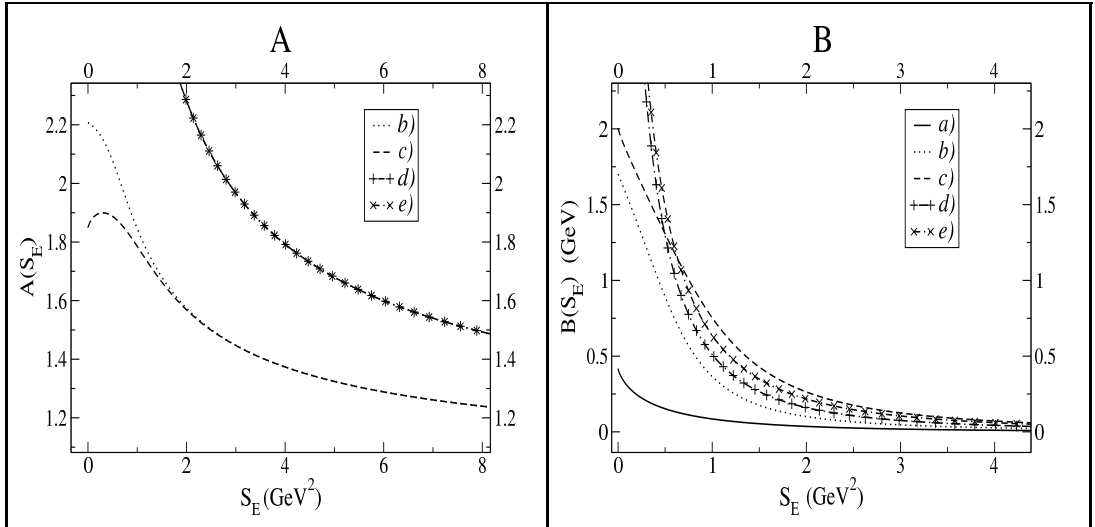,width=412pt,height=200pt}
\caption{Solutions to the \sde\ using set b in the space-like 
region. {\em a)}  Result Eq.(14) from Ref[26], {\em b)}
Eq.(26), {\em c)} Eq.(27), {\em d)} Eq.(28),
(e) Eq.(29). In (a) $A=1$.} 
\label{SLsetb}
\end{center}
\end{figure}

Of interest is also the behavior of the function 
$D\left(S_E\right) \equiv S_E A^2\left(S_E\right)+B^2\left(S_E\right)$, 
the denominator of the propagator. The vanishing of this function would 
indicate a pole in the quark propagator. Extrapolation of the behavior in 
the space-like region would suggest a pole at much smaller time-like energies 
than those we have plotted, particularly for set a (for set b some turning 
around is already noticeable at very low space-like energies). The behavior of 
$D$ right around its zero, if there is one, would be difficult to extract, 
since the functions $A \mbox{ and } B$ become singular there. In fact, 
$D$ seems to drop fast just to the left of the regions we have 
plotted. Thus, there seems to be a pole in the quark propagator on the 
real axis, in our polynomial model. The pole would be at $p^2 \approx 
(360 \mbox{ MeV})^2$ for set a, and $p^2 \approx (1.3 \mbox{ GeV})^2$ for 
set b. 
Further investigation of the solutions deeper into the time like region, 
is necessary before this can be ascertained. Notice that this refers to 
Eq.(\ref{noinst}) only.

Figures~\ref{SLseta} and \ref{SLsetb} compare the solutions to 
Eqs.(\ref{instout}, \ref{instin}, \ref{install}) for set a and set b, 
respectively. They include only the space-like region (Euclidean space), 
since they use the result (\ref{pob}) from~\cite{pob}, which was obtained 
in Euclidean space. The analytic continuation of result (\ref{pob}) 
into the time-like region exhibits a branch point at $p^2=0$ and 
thus has not been used.

\section{Conclusions.}\label{concl}

\hspace{3mm}
We have shown how to obtain a light-cone form of the \sde\ for the quark
propagator. For QCD the key question is the infra-red behavior of the
gluon propagator, for which we have used two models. Knowing that the
'tHooft model\cite{tHooft} gives confinement, we have used polynomial
models for the gluon propagator based on that model. Although the instanton
liquid form\cite{shu} for the gluon propagator does not confine, and therefore
does not have the correct far infra-red behavior, it provides the main 
mid-range QCD interaction, and allows fits to the condensates, which is
essential for obtaining hadronic properties. With our approach we can
obtain solutions for both the enhanced infra-red behavior of the polynomial
type models or the regular infra-red behavior of the instanton model and
recent lattice gauge calculations\cite{b&r93}. With the polynomial models
we are able to work in Minkowski space and obtain solutions. With the
models that include instanton effects we are able to obtain solutions that 
give much better agreement with the phenomenological values of the condensates.

We consider the present work is exploratory. It provides the framework
for obtaining light-cone QCD propagators that can be used to obtain
light-cone models of hadronic BS amplitudes for studies of hadronic
properties at all momentum transfers.

\vspace{2mm}
\hspace{3cm}
{\bf Acknowledgments}   
\vspace{2mm}

   This work was supported in part by the NSF grant PHY-00070888. The
authors would like to thank the P25 group at Los Alamos National Laboratory
for hospitality when part of this work was done, Andrew Harey and 
Montaga Aw for many helpful discussions. We would especially like to thank 
Pieter Maris for providing us with his Ph.D. thesis.

\end{document}